# Branching (Almost) Everywhere and All at Once

For the volume *Local Quantum Mechanics* (Oxford University Press, 2026)

Alyssa Ney

## 1. Introduction

It is sometimes said that one reason to prefer the many worlds interpretation of quantum mechanics (MWI) over rival realist interpretations (such as hidden variables and collapse theories) is that by adopting the MWI, one can avoid the kind of "spooky action at a distance" that is supposed to follow as a consequence of quantum entanglement according to these other approaches. It is not clear to what extent Everett himself was aware of this, although he noted in his dissertation that:

> Fictitious paradoxes like that of Einstein, Podolsky, and Rosen which are concerned with such correlated, noninteracting systems are easily investigated and clarified in the present scheme. (1957, p. 20)

This feature of the interpretation has been promoted by many of Everett's followers, including Guido Bacciagaluppi (2002), Harvey Brown and Chris Timpson (2002, 2016), David Deutsch (2012), Frank Tipler (2014), David Wallace (2012), and Lev Vaidman (2021), who notes in his *Stanford Encyclopedia* entry on the topic:

> The MWI does not have action at a distance. The most celebrated example of nonlocality of quantum mechanics given by Bell's Theorem in the context of the Einstein-Podolsky-Rosen argument cannot get off the ground in the framework of the MWI because it requires a single outcome.

But why should we think the MWI has this advantage over other realist interpretations? And how does the issue of single or multiple outcomes bear on the issue of whether there is ever interaction between distant entangled systems?

To see this, consider a standard EPR set-up, one which would generate nonlocal causal interaction according to other realist interpretations of quantum mechanics (as argued, for example, by Maudlin 2011). Alice and Bob, working in labs separated by some large distance, are each sent half of an entangled pair of particles. Suppose these particles are in the spin singlet state. We thus have a combined quantum state which we may write down as:

(1) $$\psi = \frac{1}{\sqrt{2}}|ready\rangle_A|ready\rangle_B|ready\rangle_{D_A}|ready\rangle_{D_B}|E_0\rangle_E\left(|\uparrow_z\rangle_a|\downarrow_z\rangle_b - |\downarrow_z\rangle_a|\uparrow_z\rangle_b\right)$$

Given the particles' combined quantum state, it follows from the Born rule that were Alice to measure the z-spin of her particle, she would have a 50% chance of finding it to have z-spin up and a 50% chance of finding it to have z-spin down. If Bob were to measure the z-spin of his particle, he would have a 50% of finding it to have z-spin up and a 50% chance of finding it to have z-spin down. But if both were to measure the spin of their particles, they would be guaranteed to find them to have opposite spins, assuming they both decide to measure along the same orientation.

Now let's suppose that Alice alone carries out a z-spin measurement of her particle. Bob does nothing. Alice's measurement changes the quantum state of the total system. How it changes the overall state depends on the interpretation of quantum mechanics we consider. According to the textbook collapse approach of von Neumann (1955), the state after Alice's measurement becomes one of:

(2) $$\psi = |ready\rangle_B|ready\rangle_{D_B}|\uparrow\rangle_A|\uparrow\rangle_{D_A}|\uparrow_z\rangle_a|\downarrow_z\rangle_b|E_1\rangle_E,$$

or

(3) $$\psi = |ready\rangle_B|ready\rangle_{D_B}|\downarrow\rangle_A|\downarrow\rangle_{D_A}|\downarrow_z\rangle_a|\uparrow_z\rangle_b|E_2\rangle_E.$$

And thus, it is not controversial to say that, according to a collapse interpretation, Alice's action, her measurement, has an immediate effect on Bob's particle, no matter how far away from Alice it is.[1] For it is true that the reduced density matrices associated with Bob and his measuring device have not changed as the result of Alice's measurement. Bob and his measuring device are still waiting in their "ready" states. But the reduced density matrix associated with Bob's particle *has* changed as the result of Alice's measurement.[2] It is now objectively the case that the probabilities of Bob receiving an up or down outcome, were he to measure the z-spin of his particle, have changed.

But now consider what follows if we instead adopt the MWI. Here again the total system starts in state (1). Suppose again that Alice measures the z-spin of her particle, but

---

[1] There is no great change if we instead consider the GRW theory. According to that theory, when Alice measures her particle, it is overwhelmingly likely that the system will collapse to a state roughly like:

$$\psi = |ready\rangle_B |ready\rangle_{D_B} \left[a|\uparrow\rangle_A |\uparrow\rangle_{D_A} |\uparrow_z\rangle_a |\downarrow_z\rangle_b |E_1\rangle_E + b|\downarrow\rangle_A |\downarrow\rangle_{D_A} |\downarrow_z\rangle_a |\uparrow_z\rangle_b |E_2\rangle_E\right],$$

or:

$$\psi = |ready\rangle_B |ready\rangle_{D_B} \left[b|\uparrow\rangle_A |\uparrow\rangle_{D_A} |\uparrow_z\rangle_a |\downarrow_z\rangle_b |E_1\rangle_E + a|\downarrow\rangle_A |\downarrow\rangle_{D_A} |\downarrow_z\rangle_a |\uparrow_z\rangle_b |E_2\rangle_E\right].$$

where $|a| \gg |b|$, and $a^2 + b^2 = 1$. As in the von Neumann theory, the state of Bob's particle is changed as the result of Alice's measurement. The state of Bob and his detector are unchanged.

[2] If we make the plausible assumption that the intrinsic state of a localized quantum system is given by its reduced density matrix (as in, e.g. Wallace and Timpson 2010), then it follows that the intrinsic state of Bob's particle has immediately changed as a result of this wave function collapse.

Bob does nothing. According to the MWI, the total state will then evolve not to (2) or (3), but rather instead to:

(4) $$\psi = \frac{1}{\sqrt{2}}|ready\rangle_B|ready\rangle_{D_B}\left(|\uparrow\rangle_A|\uparrow\rangle_{D_A}|\uparrow_z\rangle_a|\downarrow_z\rangle_b|E_1\rangle_E - |\downarrow\rangle_A|\downarrow\rangle_{D_A}|\downarrow_z\rangle_a|\uparrow_z\rangle_b|E_2\rangle_E\right)$$

Alice's interaction with her part of the entangled spin pair causes the universe to branch into two worlds: one in which she has a successor who finds an "up" result, another in which her successor finds a "down" result. But, crucially, as a result of Alice's measurement, there is no immediate change in the reduced density matrix associated with anything at Bob's location: not Bob, not his measuring device, not his particle. So, we should infer that according to the MWI, Alice's measurement will produce no immediate change to anything at Bob's location.[3] What is most relevant, when we are comparing the MWI to other realist interpretations of quantum mechanics is the case of Bob's particle. Prior to Alice's measurement, it was in an equal superposition of z-spin up and down. And according to the MWI, after Alice's measurement, it stays that way. (We can connect this back to Vaidman's point. Both "outcomes" are still available for Bob's particle, even after Alice has carried out her measurement.) Although there is spooky action at a distance according to other non-Everettian solutions to the measurement problem, due to the fact that Alice's measurement forces an immediate change in the state of Bob's particle, there is no such action at a distance according to the MWI.

This is a nice story.[4] However, some have argued, there is a reason to think we have not ruled out ways for non-locality to enter the MWI. For, recall that the MWI proponent will

---

[3] We can also compare with the non-local effects that Alice's measurement would have on distant systems according to hidden variables theories. Since the MWI contains no hidden variables, there is no nonlocal action in this sense either.

[4] See Ney (2025) for a detailed defense of this story.

say that when Alice does her measurement, she causes the universe to branch. And this is not a part of the interpretation that is easily dispensed with. For, in order for the MWI to provide an adequate solution to the measurement problem, measurements must lead to definite outcomes.[5] Since there is no collapse of the wave function nor variables postulated beyond the wave function that could be tracking such definite outcomes, the many worlds theorist must find these definite outcomes in total quantum states that look like (4). Of course, following Everett, the idea is that a state like (4) contains *multiple* definite outcomes: one in which there is a successor of Alice who finds "up," and one in which she finds "down." But to say these outcomes are *definite*, one requires more than just that one can read these outcomes off of mathematical formulations like (4). The standard story today is that for (4) to describe multiple, definite outcomes, (4) must represent a decoherent quantum state and this is to ensure that there is very little if any causal interference between the part of the universe in which Alice sees an up result and the one in which Alice sees a down result. The outcomes each Alice finds must be unambiguous. As we will discuss in more detail below, this is why, to solve the measurement problem, the MWI proponent typically says that measurements lead to branching. In most versions of the MWI today, branching is understood *just as* this event or process by which parts of a total quantum system – here, "Alice's seeing up" and "Alice's seeing down" – become (for the most part) causally isolated. It is thus branching that enables

[5] As Maudlin (1995) correctly notes, there are many "measurement problems." Here I have in mind the measurement problem as presented in Albert (1992). This is the problem of reconciling the fact (a) that when one measures a system in a superposition of some observable to determine its value with respect to that observable, the Schrödinger equation implies it must stay in a superposition of that observable, with the fact (b) that such measurements, when properly conducted, have definite outcomes.

the MWI to secure definite measurement outcomes, and thus a solution to the measurement problem.[6]

So, according to the MWI, when Alice conducts her measurement, her intervention causes the universe to branch by way of the different parts of the total quantum system becoming (to a large extent) causally isolated. After the measurement then, there are two Alices, one recording a definite "up" outcome, and one a definite "down" outcome. But now we may ask: if it is true that the *universe* branches as the result of Alice's measurement, then doesn't this mean that everything described in (4) has undergone a branching? But if this is the case, then it does seem that there is spooky action at a distance according to the MWI. For the measurement that triggered a splitting of Alice into two copies that largely and for the most part cannot interact with each other must have at the same time caused a splitting of Bob into copies that also mostly cannot interact with each other. (And likewise, for Bob's measuring device, his particle, and everything else in the universe.) And given that branching events are taking place all of the time, and always immediately affect objects very distant from the location where the branching was initially triggered, it would seem that nonlocality is utterly ubiquitous; it is not avoided at all on the MWI.

[6] This general claim needs a comment. Actually, not every version of the MWI tries to achieve the causal separation that is required for definite outcomes by the postulation of branching. In particular, Alistair Wilson (2020) and Isaac Wilhelm (2022) propose versions of the MWI in which the two Alice worlds (e.g.) are not *generated* by a branching event, but rather are always there and numerically distinct, even before Alice's measurement is conducted. Wilson uses language he inherits from David Lewis (1986) in order to distinguish his view (which is in this respect more like Lewis's) from the more common one associated with the MWI today: for Wilson, the Everettian multiverse has a *divergence* rather than a *branching* structure.

What is the MWI proponent to say? There are various responses available. Some MWI proponents concede the point, saying "yes," there is this kind of action at a distance and it is an unavoidable part of the MWI (Sebens and Carroll 2018, Carroll 2019).[7] More often however, MWI proponents attempt to show that these appearances are deceiving, and that if we properly understand branching in the MWI, we will see how it does not involve any action at a distance. My aim in this chapter is to evaluate one way of arguing that branching does not involve any action at a distance. This way is based on a proposal that was first advocated explicitly by Wallace (2012), but has more recently been developed by Nadia Blackshaw, Nick Huggett, and James Ladyman (this volume). I will call this *the local branching view*. [8] Wallace argues that if we properly understand the nature of branching, especially in a relativistic quantum universe, we will understand that it is a local causal process, one that starts in a localized spacetime region and only spreads throughout the universe at light speed or subluminally. If one accepts the local branching view, then one can diagnose the problem with the above argument to be with its assumption that branching occurs instantaneously and everywhere across an entire time-slice of the universe. If branching is rather a local causal process, then it does not affect everything in the universe instantaneously. And so, Alice's measurement does not instantaneously cause Bob or anything at his location to branch. Alice's measurement can lead to Bob's branching, but, to put it somewhat technically, this will occur no earlier along Bob's worldline than at the location his worldline intersects the lightcone from Alice's measurement.

---

[7] Sebens and Carroll (2018) call the resulting action at a distance "psychologically unintuitive but empirically benign" (p. 35).

[8] To my knowledge, it was Sebens and Carroll (2018) who first made the distinction between 'local' and 'global' branching views explicit.

At first glance, this is a promising way of filling out the MWI. It makes the branching process appealingly relativistic. Moreover, although the MWI needed something like a branching process in order to show how quantum states like (4) are compatible with the existence of definite outcomes and solve the measurement problem, it isn't clear why this should have required that branching take place instantaneously over an entire slice of the universe. If Alice and Alice alone measures her particle, then we just need a story according to which *she* sees a definite result. So, she must branch. But there is no immediate reason why Bob must branch as well, at least at that time. If at some time later, Alice goes on to share her measurement result with Bob, then he will have to branch too in order that he (or better, his successors) see definite outcomes. But Bob's branching need not happen immediately after Alice's measurement. After all, it will take some time before Alice is able to communicate her result with Bob. In conclusion, if Alice's measurement only immediately causes herself and objects in her vicinity to branch, and not distant objects like Bob or his particle, then it seems the MWI can solve the measurement problem and at the same time avoid spooky action at a distance in the form of superluminal influence.

My aim in this paper is to show, however, why we should *not* adopt the local branching view. I will not in any way contest the claim that if we were to adopt the local branching view, then this would give us a straightforward way to avoid spooky action at a distance; I think it would. My claims are rather that: (1) we don't need branching to be local to avoid spooky action at a distance, and (2) that the right way to conceive of branching in the MWI takes it to be a global and instantaneous event and not a local, causal process.[9] Most of the remainder of this chapter will be concerned with arguing for (2). In the penultimate section, I turn back to (1) and show how the global branching view is unproblematic from the point of view of relativistic quantum theories and does not involve any spooky action at a distance.

## 2. The Case for Local Branching

So, according to what we may call *the global branching view*, when Alice conducts her measurement, objects across the entire universe branch. This implies that if at some time t, Alice branches, then, if Bob exists at t, then he also branches at t. The global branching view is a natural way to interpret the quantum formalism, at least in the way we have presented it here, using the Schrödinger picture.[10] After all, recall the way we represented the state resulting from Alice's measurement of her particle:

(4) $$\psi = \frac{1}{\sqrt{2}}|ready\rangle_B|ready\rangle_{D_B}\left(|\uparrow\rangle_A|\uparrow\rangle_{D_A}|\uparrow_z\rangle_a|\downarrow_z\rangle_b|E_1\rangle_E - |\downarrow\rangle_A|\downarrow\rangle_{D_A}|\downarrow_z\rangle_a|\uparrow_z\rangle_b|E_2\rangle_E\right)$$

---

[9] I should be explicit that my aim in this chapter is really just to clarify the way that advocates of the MWI should interpret the nature of branching and how this relates to the MWI's avoidance of spooky action at a distance. My discussion especially of the latter issue will likely not convince most MWI-skeptics. For example, Travis Norsen (2016) has argued that there is unavoidable non-locality in the MWI, but his argument relies on the claim that the MWI can include no satisfactory account of the probabilities in the Born rule. As we will see, I agree with most proponents of MWI that there are actually several ways to make sense of the probabilities in the Born rule on the MWI. But I recognize that MWI-skeptics disagree, and indeed this is often the primary reason they claim to be MWI-skeptics.

[10] I say 'natural' here, not 'obligatory.'

This is of course mathematically equivalent to:

(5) $$\psi = \frac{1}{\sqrt{2}}\Big(|ready\rangle_B |ready\rangle_{D_B} |\uparrow\rangle_A |\uparrow\rangle_{D_A} |\uparrow_z\rangle_a |\downarrow_z\rangle_b |E_1\rangle_E - |ready\rangle_B |ready\rangle_{D_B} |\downarrow\rangle_A |\downarrow\rangle_{D_A} |\downarrow_z\rangle_a |\uparrow_z\rangle_b |E_2\rangle_E\Big),$$

which appears to be a perspicuous representation of two worlds containing not just two Alices, but also two Bobs.[11] As we have seen, proponents of the local branching view disagree. According to *the local branching view*, when Alice conducts her measurement, the state of the universe evolves to (4), but only the objects at Alice's location branch; the impression conveyed by (5) that Bob has also branched is deceiving.[12]

There are two (related) points of disagreement between the global branching view and the local branching view. First, there is a disagreement about where in spacetime branching takes place (at least initially): across an entire timeslice or spacelike hypersurface, or at some far more localized region. Second, there is a disagreement about whether branching is an instantaneous event (defined along a spacelike hypersurface) or a causal process that unfolds (and so is more accurately defined along a timelike or lightlike hypersurface). Carroll and Sebens (2018), Carroll (2019), Vaidman (1998), and McQueen and Vaidman (2019) all advocate the global branching view. Wallace (2012) and Blackshaw, Huggett, and Ladyman (this volume) advocate the local branching view. I find other work on the MWI rather more ambiguous in discussions of how to understand branching. The view presented in Deutsch (2012), Deutsch and Hayden (1999), and Kuypers and Deutsch (2021) is similar in key respects to the local branching view. However, although what is described in those papers is clearly a version of Everettian quantum mechanics in being a realist picture that rejects both objective collapse of the wave function and hidden variables, it is not a many worlds theory (see Timpson 2004). Although there is a local causal process in this model that is something like local branching, it is not branching in the sense discussed in the present chapter.

[11] See also Sebens and Carroll (2018), p. 34.

[12] This raises the question of whether one who adopts the local branching view ought to think we would do better to adopt a quantum formalism that more accurately tracks the branching structure (e.g. something like the Deutsch and Hayden (1999) formalism). My sense is that those who adopt the local branching view do not think this, and this is because they do not regard the branching process as a fundamental physical process that needs to be perspicuously represented by the formalism.

As I mentioned in the previous section, an initial reason in favor of the local branching view is the desire to make all processes compatible with special relativity, and so causal. This is a motivation we would do well to call into question. After all, although there is certainly good reason to take all fundamental physical processes to be causal, it is not clear that branching itself is correctly regarded as a fundamental physical process. Branching is the phenomenon that generates worlds. And worlds, after all, are emergent ontology according to most MWI proponents today.[13] To say they are emergent doesn't undercut their reality (Wallace 2010). But it does undercut the requirement that the way they are generated must be by way of a causal process. To use an analogy, consider in-laws. In-laws are real entities. However, they are emergent. Thus, I take there to be no reason to think that the process that generates in-laws must be a causal process. If I marry your brother while you are far away, I don't need to wait until the lightcone from the ceremony intersects your worldline for me to correctly say that you are now my in-law.[14]

---

[13] "The number of worlds is not a physical parameter in the theory," (Vaidman 1998, p. 13).

[14] Analogies using social construction are easy to state and make the point, however one might worry that branching isn't a social phenomenon and so the analogy does not hold up. I continue to use the social analogy in what follows, but note that non-social analogies are

This is not to say that the best motivation to see branching as a local causal process comes from the belief that it is a fundamental physical process, and so branching must, for that reason, be causal. Rather I think the main motivation to see branching as a local causal process is that most proponents of the MWI today regard branching as the result of decoherence, and decoherence is a local causal process. Thus, there appears to be a simple two-premise argument in favor of the local branching view:

1. Branching is the result of decoherence.

2. Decoherence is a local causal process.

Therefore,

3. Branching is a local causal process.

Both of these premises are very compelling on their own. It is thus understandable that some proponents of the MWI would adopt the local branching view. Let's take a moment then to unpack each of these two premises.

We may see the first premise of this argument as motivated by a key component of the MWI agreed to by many of its proponents today. This is the functionalist analysis of worlds. The functionalist analysis of worlds comes out of the way that many of today's proponents of the MWI solve the measurement problem – i.e. the way that they reconcile the sort of states that result from the measurement of systems in a superposition of the observable that is measured (such as (4)) with the definite outcomes we know that observers routinely find

---

readily available. For instance, suppose a new kind of plant evolves that triggers the sort of immune response in humans that are associated with peanut allergies. Then it will at that time be instantaneously true that humans, wherever they live, have this allergy. Like being someone's worldmate (or not), having an allergy is a dispositional feature defined by causal relationships that could or could not obtain in certain situations.

when they carry out such measurements. As discussed above, "worlds" is the label used in the MWI to refer to the subsystems of the total universe in which definite outcomes may be found; subsystems which, to achieve definiteness, must be (mostly) causally isolated from one another. Along these lines, Wallace proposes the following functionalist analysis of worlds:

> "Worlds" are mutually dynamically isolated structures instantiated within the quantum state, which are structurally and dynamically "quasiclassical." (Wallace 2010, p. 17)

This is an *analysis* in the sense that it gives truth conditions for claims about the existence of multiple worlds. Despite the appearance of the scare quotes, Wallace is insistent that these worlds exist according to the MWI in as literal a sense as there can be. This is a *functionalist* analysis in the sense that the conditions required for the existence of worlds are causal or dynamical conditions. They specify what kind of behavior in the more basic ontology (the quantum state evolving according to unitary dynamics) is sufficient to make it the case that there exist two or more worlds.

This is an account of "worlds" that has a history in both metaphysics and cosmology. It is considered, though rejected, by Lewis (1986). It is also discussed by Max Tegmark (2014) in the context of his several multiverse models (e.g. his Level I and II multiverses), not only the MWI (his Level III multiverse). The way this functionalist analysis is developed by Simon Saunders (1993), Vaidman (1998), Wallace (2010, 2012), and Carroll (2019) is that what makes it true that there are multiple worlds rather than one single world for the MWI proponent is not, as for Lewis, that there exist numerically distinct spatiotemporally connected systems that are in turn *spatiotemporally* isolated from one another. Rather, it is that there are parts of the total quantum state (a quantum state that could be describing one single connected spacetime) that are (for the most part) *causally* isolated from one another. In virtue of this high degree of causal isolation, these subsystems (largely and for the most part) evolve independently of one another. And this allows them to be described as quasi-classical systems like particles with (fairly) definite locations and properties, measuring devices with pointers pointing in clear directions, and people with determinate beliefs. The part of the total system in which Alice believes the pointer is pointing up does not noticeably affect the part in which Alice believes the pointer is pointing down, and this is why each Alice is truly said to have received a definite outcome for her measurement.

The functionalist analysis of worlds is the first component of the justification for premise (1). The second component is the meaning of decoherence. Decoherence is the process whereby parts of the total quantum state become largely and for the most part causally isolated from one another. When we have a coherent quantum state, we have a state in which there is interference. A system in a coherent state will have a density operator that contains the presence of significant interaction (or "cross") terms that track the interaction between the parts of the quantum state. For example, consider a situation in which Alice alone has been given a particle in a superposition of z-spin states to measure. Suppose the wave function of the total system may be written as:

(6) $$\psi = \frac{1}{\sqrt{2}}|ready\rangle_A|ready\rangle_D|\uparrow_z\rangle_P|E_0\rangle_E + \frac{1}{\sqrt{2}}|ready\rangle_A|ready\rangle_D|\downarrow_z\rangle_P|E_o\rangle_E$$

Taking the outer product to arrive at the density matrix and tracing out the environment, we arrive at the reduced density matrix:

(7) $$\rho = \frac{1}{2}|ready\rangle_A|ready\rangle_D|\uparrow_z\rangle_P\langle ready|_A\langle ready|_D\langle\uparrow_z|_P +$$

$$\frac{1}{2}|ready\rangle_A|ready\rangle_D|\uparrow_z\rangle_P\langle ready|_A\langle ready|_D\langle\downarrow_z|_P +$$

$$\frac{1}{2}|ready\rangle_A|ready\rangle_D|\downarrow_z\rangle_P\langle ready|_A\langle ready|_D\langle\uparrow_z|_P +$$

$$\frac{1}{2}|ready\rangle_A|ready\rangle_D|\downarrow_z\rangle_P\langle ready|_A\langle ready|_D\langle\downarrow_z|_P$$

Decoherence is a suppression of this interference. When Alice now carries out the measurement of her particle, the quantum state evolves (for reasons to be discussed) to:

(8) $\psi = \frac{1}{\sqrt{2}}|\uparrow\rangle_A|\uparrow\rangle_D|\uparrow_z\rangle_P|E_1\rangle_E + \frac{1}{\sqrt{2}}|\downarrow\rangle_A|\downarrow\rangle_D|\downarrow_z\rangle_P|E_2\rangle_E$

This state is such that when we take the outer product to arrive at the density matrix and trace out the environment this time, we arrive at the reduced density matrix:

(9) $\rho = \frac{1}{2}|\uparrow\rangle_A|\uparrow\rangle_D|\uparrow_z\rangle_P\langle\uparrow|_A\langle\uparrow|_D\langle\uparrow_z|_P + \frac{1}{2}|\downarrow\rangle_A|\downarrow\rangle_D|\downarrow_z\rangle_P\langle\downarrow|_A\langle\downarrow|_D\langle\downarrow_z|_P$

In effect, the result of a quantum system in a microscopic superposition's interaction with a measuring device, observer, and larger environment leads to a quantum system that is accurately, if approximately, described as one in which there are a multiplicity of systems with determinate locations that do not interfere or interact with one another.[15] In short, in situations like the measurements of systems in quantum superpositions, we have a physical process of decoherence in the overall system that makes it the case that there are parts of the total quantum system correctly described as many worlds. Thus, decoherence is what allows a branching of a single universe into multiple worlds. The first premise, then, is analytic. It follows from the functionalist analysis of worlds and the meaning of a decoherent versus coherent quantum state.

The second premise follows from contingent facts about how this decoherence is achieved in practice. As Wojciech Zurek (2003) explains, it is a process that unfolds over time. The kind of decoherence we have just been discussing is environment-induced superselection (Schlosshauer 2007), where an event leaves traces in its environment and this leads to a suppression of interference. Return to Alice's measurement of the z-spin of her particle as illustration. Her particle interacts with her measuring device and causes a pointer to move to a certain position on the detector screen. She sees the pointer and makes a mark in her notebook. Light reflects off her pen as she writes. As the two parts of the quantum state continue to leave distinct traces, the environmental components of the total state become more and more distinct. And as the environmental states associated with the two terms become increasingly orthogonal, this suppresses the cross- or interaction terms underwriting quantum coherence. Thus, decoherence is achieved in practice by processes involving signals traveling at or below light speed, starting at a source and then spreading outward. The signals will reach the objects in Bob's environment no sooner than the point when these objects' worldlines intersect the light cone spreading out from Alice's measurement.

[15] (8) and (9) are idealized representations of real systems. In practice, the cross-terms will not be eliminated when Alice carries out her measurement, only largely and for the most part suppressed, as indicated by a decrease in their coefficients.

Putting these two premises together – branching is the result of decoherence and decoherence is a local causal process – it is natural to draw the conclusion that branching is also a local causal process. As Wallace puts it:

> When some microscopic superposition is magnified up to macroscopic scales (by quantum measurement or by natural processes) it leads to a branching event which propagates outwards at the speed of whatever dynamical interaction is causing decoherence – in practice, it propagates out at the speed of light. (Wallace 2012, p. 307)

Blackshaw, Huggett, and Ladyman agree:

> [Our] model exemplifies how we have given up the idea that the whole world branches instantaneously on measurement; rather it splits into extended branches that grow over time. (Blackshaw, Huggett, and Ladyman this volume, p. 15)

So then, why would an advocate of the MWI choose to instead adopt the global branching view? I will consider four reasons.

## 3. Four Objections to the Local Branching View

### *3.1 Branching is Not the Result of Decoherence*

The first reason why an advocate of the MWI might object to the view that branching is a local causal process is that they reject the first premise of the argument of Section 2. They reject the premise that branching is the result of decoherence. As we saw that this premise analytically follows from (a) the functionalist analysis of worlds and (b) the meaning of 'decoherent state,' one must then reject either (a) or (b). Likely, one will reject the more philosophically-motivated (a) and offer instead some other analysis of what it means to say there exist one or many worlds. Vaidman has been most explicit among advocates of the MWI that he does not take branching to be the result of decoherence. This resistance indeed can be traced to his rejection of the functionalist analysis. According to Vaidman, "worlds are subjective concepts of observers" (1998, p. 13). This does not mean that Vaidman rejects the reality of environment-induced superselection; he just does not take this process to be constitutive of branching.

In what follows, I will not pursue this way of rejecting the local branching view. Like most other advocates of the MWI, I find both premises of the argument of Section 2 compelling. Although skepticism about the functionalist analysis of worlds and thus the link between decoherence and branching is a consistent way to cast doubt on the motivation for the local branching view, it is not a way I will pursue in this chapter. It is also worth noting that this is not the only reason Vaidman has to reject the local branching view. We will discuss another reason shortly.

*3.2 Violation of Quantum Statistics?*

The second concern is that one might worry that the adoption of the local branching view leads to a violation of the (observed) quantum statistics. This concern turns out to rest on a simple confusion, but it is instructive to see why, as it will allow us to see the local branching view more clearly.

To understand the concern, let's take a brief detour and consider again rival collapse interpretations of quantum mechanics. We have been focused on the way that an advocate of the MWI might argue there is no spooky action at a distance in EPR cases because the branching process is local. But we might ask: why couldn't the collapse theorist also avoid spooky action at a distance, by adopting the view that the collapse process is local? This collapse theorist could make a claim analogous to that we just considered from Wallace, saying:

> When some microscopic superposition is magnified up to macroscopic scales (by quantum measurement or by natural processes) it leads to a *collapse* event which propagates outwards at the speed of whatever dynamical interaction is causing decoherence – in practice, it propagates out at the speed of light.

The problem is that adoption of this view would involve a simple violation of quantum statistics.[16] Here is why.

Let's suppose that Alice conducts her measurement, and this leads to a collapse of the wave function. But now since we are adopting a *local* collapse theory, we say that as the result of Alice's measurement, only the parts of the wave function describing what is in the vicinity of Alice collapse immediately. The parts of the wave function describing what is at Bob's location will not collapse any sooner than the points on these objects' worldlines intersect the light cone coming from Alice's measurement.

In mathematical terms, let's assume the quantum state we started with again is:

(1) $$\psi = \frac{1}{\sqrt{2}}|ready\rangle_A|ready\rangle_B|ready\rangle_{D_A}|ready\rangle_{D_B}|E_0\rangle_E\left(|\uparrow_z\rangle_a|\downarrow_z\rangle_b - |\downarrow_z\rangle_a|\uparrow_z\rangle_b\right).$$

[16] Thanks to Emily Adlam for discussion.

And let's suppose that Alice receives an "up" result from her measurement. Then the view we are considering is that we avoid spooky action at a distance because state (1) does not evolve to:

(2) $$\psi = |ready\rangle_B |ready\rangle_{D_B} |\uparrow\rangle_A |\uparrow\rangle_{D_A} |\uparrow_z\rangle_a |\downarrow_z\rangle_b |E_1\rangle_E,$$

like the textbook collapse theory claims. Instead, (1) evolves to:

(10) $$\psi = \frac{1}{\sqrt{2}} |ready\rangle_B |ready\rangle_{D_B} |\uparrow\rangle_A |\uparrow\rangle_{D_A} |\uparrow_z\rangle_a \left[ |\downarrow_z\rangle_b |E_2\rangle_E + |\uparrow_z\rangle_b |E_3\rangle_E \right],$$

and the collapse process takes some time to reach all of the way to where Bob is. But now we can see the problem. If (10) is the quantum state that results from Alice measuring her particle, then it follows from the Born rule that even though Alice has already found her particle to be z-spin up, there is a 50% chance Bob will also find his particle to be z-spin up. But we know from the many tests that have experimentally confirmed quantum mechanics that in circumstances like this, there is zero chance that Bob will measure z-spin up. This would violate the conservation of total spin. In conclusion, adopting a local *collapse* theory is a very bad idea indeed. This would imply a violation of the observed quantum statistics.

Is there a similar problem if one instead adopts the local *branching* view? It is straightforward to see that the answer is "No." The key difference is that, as we have just seen, to adopt a local collapse view is to adopt a different quantum theory than "global" collapse theory, one that makes substantially different experimental predictions from the standard theory, predictions that turn out to be wrong. On the other hand, whether we adopt a local or global branching view, we will regard the evolution of the quantum state in the same way. Before Alice conducts her measurement, the quantum state is (1). After she conducts her measurement, the quantum state is:

(4) $$\psi = \frac{1}{\sqrt{2}} |ready\rangle_B |ready\rangle_{D_B} \left( |\uparrow\rangle_A |\uparrow\rangle_{D_A} |\uparrow_z\rangle_a |\downarrow_z\rangle_b |E_1\rangle_E - |\downarrow\rangle_A |\downarrow\rangle_{D_A} |\downarrow_z\rangle_a |\uparrow_z\rangle_b |E_2\rangle_E \right)$$

Both before and after Alice conducts her measurement, it follows from the quantum state and Born rule that there is no chance that both Alice and Bob will find their particles to have z-spin up. There is no world in which this occurs. So there is no violation of the observed quantum statistics on the local branching view.

*3.3 Tension with Accounts of Probability in MWI*

A third reason one might be skeptical of the local branching view has to do with the way this model of branching interacts with one's account of probability in the MWI. It is frequently argued that the MWI is inconsistent with the probabilistic claims made by the Born rule, such as that given a quantum state $\psi$, the probability one will find a system to have value *a* is $|\langle a|\psi\rangle|^2$. According to the MWI, all laws of temporal evolution are deterministic. And, according to the MWI, there is nothing in the world that is not described by the quantum state. So, once one knows the quantum state, all facts about the measurement outcomes that will follow are determined. To continue with our example, if Alice knows the initial quantum state is:

(1) $$\psi = \frac{1}{\sqrt{2}}|ready\rangle_A|ready\rangle_B|ready\rangle_{D_A}|ready\rangle_{D_B}|E_0\rangle_E\left(|\uparrow_z\rangle_a|\downarrow_z\rangle_b - |\downarrow_z\rangle_a|\uparrow_z\rangle_b\right),$$

she can already know at the time she prepares her experiment what will follow if she measures her particle. Her measurement will trigger decoherence in the quantum state such that the system will undergo branching into two worlds. In one of these worlds a successor of Alice will observe an up result. In the other, a successor of Alice will observe a down result. Given all of this, it is reasonable to complain, as Albert (2010) and other critics of the MWI have, that this makes the MWI conflict with the Born rule. For it looks like it follows from the MWI that in a state like (1), the probability that a future Alice will find an up outcome is 100%. And also the probability that a future Alice will find a down outcome is 100%.

All proponents of the MWI disagree with Albert on this point, however they have different ways of making sense of the probabilities in the Born rule. One idea that several MWI proponents are sympathetic to (Vaidman (1998), Sebens and Carroll (2018), McQueen and Vaidman (2019), Carroll (2019)) is that we should interpret the probabilities in the Born rule as measures of self-locating uncertainty. Here is the idea. True, before Alice conducts her measurement, there is nothing for her to be uncertain of. However, if we suppose she conducts her measurement, and so she branches, but she hasn't yet looked at the result of her measurement, then there is something she can be uncertain of: which world she is in. And so, in this scenario it can be reasonable for Alice to assign a probability of 50% to the proposition that she is in the world in which her particle is z-spin up and a probability of 50% to the proposition that she is in the world in which her particle is z-spin down. Here is Sebens and Carroll:

> Thus even if she (incredibly) knows the universal wave function exactly, Alice still has something to be uncertain of. She isn't uncertain about the way the universe is; by supposition, she knows the wave function and this gives a complete specification of the state of the universe. Alice is uncertain about where she is in the quantum multiverse … She doesn't know if she's in the branch of the wave function in which the detector displays up or the one in which it shows down. We say that Alice has 'self-locating uncertainty' … We call this period in which self-locating uncertainty is present, after the measurement has been made and branching has occurred via decoherence but before the experimenter has registered the result, the 'post-measurement, pre-observation' period. (2018, p. 35)

There has been a lot of discussion of whether this proposal works, and if so, how exactly it should be formulated to make it work. We need to know how to get the probabilities to match the Born rule at all times, not just in these highly contrived circumstances in which someone has done a measurement, but hasn't yet looked at the result. We need to know how to get probabilities with values that correspond exactly to the Born rule, rather than what may be suggested by using simple principles of indifference. But this is not our concern here. Rather, our concern is whether the self-locating uncertainty account of probability in the MWI requires a certain model of the branching process.

There is a reason to think that it does, and that in particular, the self-locating uncertainty account of the probabilities in the MWI relies on the global branching view. Here is why. Suppose again that Alice conducts her measurement and the quantum state evolves to (4). And let's assume the good case for the self-locating uncertainty account, where the Born rule is valid because although Alice has branched, she doesn't yet know which world she is in. If the Born rule is valid, then this also means that there is a 50% chance that Bob's particle will be found z-spin up and a 50% chance that Bob's particle will be found z-spin down. But that requires that in each of the two worlds that are compatible with Alice's beliefs up to this time, there are two "Bob-particles," particles that we have already stipulated are located at a great distance away from Alice. For this account of the Born rule to work, Alice must be uncertain about whether she is in the world with the particle that would be found down if its z-spin were measured, or in the world with the particle that would be found up if its z-spin were measured. So, at the time Alice has branched and has the sort of self-locating uncertainty that makes the Born rule valid, objects at Bob's location must have branched. And this is so no matter how far away these objects are. Thus, the self-locating uncertainty model of probability in the MWI appears to require the global branching view. Thus, if one thinks that the correct way to make sense of probability in the MWI is in terms of self-locating uncertainties, then one has a reason to prefer the global branching view.[17]

The key word here is 'if.' Notably, none of those who advocate the local branching view to my knowledge have advocated we adopt the self-locating uncertainty account of probability in the MWI. Wallace (2012) advocates a very different account, which derives the Born rule instead from facts about what sort of bets an agent, including one facing a branching situation, would take or decline.[18] There is no reason to think objects must have already branched in order for certain betting decisions to be reasonable or unreasonable. Thus, this view about probability has no implications about whether branching is global or local. The same may straightforwardly be said about the branch counting proposal of, for example, Saunders (forthcoming).

The upshot is that MWI proponents who adopt the self-locating uncertainty account of the Born rule like Vaidman, McQueen, Sebens, and Carroll have an interesting package deal, one that includes the global branching view. But unless we have some way to argue that we should prefer this package to the others (in particular, to Wallace's decision-theoretic account of probability + the local branching view), we don't yet have a dialectically compelling reason to reject the local branching view.

---

[17] This is not an argument that has been advocated previously by proponents of the self-locating uncertainty account of probability in MWI. Sebens and Carroll (2018) do reject a kind of local branching view, but the reason that they give is that local branching would rely on an "arbitrary" carving of the universe into subsystems (p. 34). If I understand the concern correctly, I do not think it applies to the way that local branching is modeled by Blackshaw, Huggett, and Ladyman, which is very much in the spirit of what was advocated by Wallace. The carving into subsystems is just a carving into spacetime regions. Of course, it is arbitrary which level of grain one applies in describing spacetime regions and so these authors' local branching process, but this is not a problem. The local branching proponent need not claim one such carving is ontologically privileged.

[18] This kind of decision-theoretic account was initially proposed by Deutsch (1999).

### *3.4 Tension with the Functionalist Analysis of Worlds*

So far, I have discussed three (maybe four, if you count footnote 17) reasons to be skeptical of the local branching view. I have explained why I don't find any of these reasons compelling. However, there is one last reason I will consider for why one might be skeptical of local branching, and this, I claim, is a good reason. It is that the local branching view is in tension with the functionalist analysis of worlds. Recall:

> *The functionalist analysis of worlds*: worlds are parts of the total quantum state that behave for the most part like causally isolated, independently evolving, quasi-classical systems.

Fix Alice's rest frame as the reference frame for this discussion. And now consider *the Alice who measures z-spin up* that is described by (4). The key question is: at this time t when Alice has branched, are there two Bobs, one who is (mostly) causally isolated from this Alice and one who is not? If yes, then the local branching view is false.

I propose the following argument to establish this conclusion:

1. If the Born rule is valid, then, given the quantum state at t, counterfactuals such as (CF) below are true for the Alice who sees up at time t in her frame.
2. But if counterfactuals like (CF) are true for this Alice at t, and counterfactuals track facts about causal dispositions, then the following claim about causal isolation (CI) follows, which implies that this Alice is causally isolated from one Bob, but not from another.

3. If (CI) obtains, then the functionalist analysis of worlds implies that Bob has already branched at t.

Therefore,

4. The functionalist analysis of worlds (combined with the Born rule and the fact that counterfactuals track facts about causal dispositions) implies that the local branching view is false.

The counterfactual whose truth for Alice I claim follows from the Born rule and the quantum state at t is:

(CF) Were some Bob to measure the z-spin of his particle, and I were to ask him to tell me the result he found, he would tell me he received a down result. I would not hear from him that he received an up result.[19]

Alice can know this counterfactual is true, even though she knows that there are two Bobs in the universe, and that one of these successors, were he to measure the z-spin of his particle, would find it to have z-spin up. Moreover, there are many other counterfactuals that Alice can know at this time about what would happen were Bob to measure the spin of his particle along other orientations besides the z-axis, and the probabilities that he would tell her he received this or that result, were she to ask. These also follow from the fact that Alice herself received an up-result, the quantum state (4), and the Born rule.

Even those who reject the claim that causal claims may be *analyzed* in terms of counterfactuals (e.g. Maudlin 2007, 2011) agree that counterfactuals are at least indicators of causal dispositional facts. Thus, I argue that counterfactuals like (CF) justify certain facts about (a) the possibility of causal interaction between Alice and a Bob who, in these counterfactual measurement situations, has certain probabilities of telling Alice he received this or that result, and (b) the lack of possible causal interaction between Alice and a Bob who, in these counterfactual measurement situations, has distinct probabilities of telling Alice he received this or that result.[20] Therefore, it follows already at t that:

> (CI) Alice is not causally isolated from the Bob who would receive a down result, if he were to measure the z-spin of his particle. Though she is (at least for the most part) causally isolated from the Bob who would receive an up result, if he were to measure the z-spin of his particle.

[19] And here, we can set aside the contrived set-up from Section 3.3, in which Alice does not know the result of her measurement. For our purposes here it doesn't really matter, but let's keep things simple and assume Alice has looked at her device, and is now at t consciously aware that she has received an up result.

And this means that as soon as she has branched, this Alice is already causally isolated from one of the Bobs, but not the other. So, at this time, by Leibniz's Law, there are already two Bobs. The falsity of the local branching view thus seems to follow from the Born rule, the fact that certain counterfactuals track facts about causal dispositions, and hence causal isolation, and the functionalist analysis of worlds.

---

[20]Admittedly, establishing causal isolation via counterfactuals is not as direct as establishing causal connectedness via counterfactuals. My claim, however, is that none of the counterfactuals that would have to be true in order for Alice to be causally connected with this other Bob obtain. For example, the counterfactual "Were Bob to measure the z-spin of his particle, and I were to ask him to tell me the result he found, he would tell me he received a up result," is false for Alice at t, even though she knows that in this scenario, there exists a Bob, at the same location as the Bob with whom she can interact, who receives an up result.

Objection: The counterfactuals just tell us what Bob would or would not do, not any categorical facts about what he is like now. Since these dispositional facts are true in virtue of what could happen later, it doesn't require that Bob has already branched at t.

Reply: But what the counterfactuals like (CF) and thus (CI) establish is that there is one Bob who at t already has one disposition, and one Bob who lacks that disposition. And that means that already at t, it is true that there are two Bobs, even if the dispositions concern what these Bobs might do or say at later times.

Objection: But couldn't we interpret the "Bob" who is being referred to in (CF) as one of the Bobs who exists sometime after the worldline of the original Bob intersects the lightcone from Alice's measurement event? In that case, then we would not need Bob's branching to have occurred already at t.

Reply: It is possible to interpret the counterfactual that way if we make some further stipulations. If we do so, then the truth of (CF) is compatible with both the local and the global branching views. But what matters is that it is also possible to interpret the counterfactual so that it refers to one of the Bobs who Alice knows to exist at t, the one she knows she is able to causally interact with. If we like, we can force this second reading by adding an additional stipulation to the case that Bob dies immediately after t. Here it might be useful to return to the in-law analogy. Suppose you were not able to attend my wedding to your brother. At the exact time the ceremony is completed, I can correctly congratulate myself on gaining a new in-law. And this claim will be correct even if you do not survive past t. In this case, it is still true that we were in-laws, if only for a moment.

The same point can be used to rebut another more subtle objection that assumes Lewis's (1976) perdurance account of persistence through cases of branching (what Lewis refers to as fission cases):

Objection: At time t, shouldn't all many worlds theorists believe that there are already two Bobs present, even defenders of the local branching view? After all, the correct metaphysics of persistence, perdurantism, implies that people are spacetime worms. Since in the future, Bob branches, this means already at time t, there are two worms, and so two Bobs that are numerically distinct. It is just that at t, these two worms overlap in virtue of sharing a common temporal part. (Similarly, they were distinct at all previous times from Bob's birth. They just overlapped in sharing common temporal parts.) Therefore, given perdurantism, we don't need to say that Bob branched already at t, in order to allow the truth of counterfactuals like (CF).

Reply: There is a lot to say here about whether perdurantism is the correct metaphysics of persistence the many worlds theorist should adopt.[21] However, the most direct response is to again to appeal to the case in which Bob dies immediately after Alice's measurement and so, according to the local branching view, would never go on to either make a measurement or receive the information from Alice's measurement. In this case, perdurantism will not itself implies that there are two overlapping Bobs at t or at any earlier time. Yet, again, I argue that the counterfactuals (CF), and their consequence (CI), by way of the functionalist analysis of worlds, will imply that already at t, there are two Bobs because, at that time, Bob has branched.

This concludes my consideration of objections. But now it might seem that something has gone wrong. After all, recall that in Section 2, we considered an argument for the local branching view which had two very well-supported premises. Since I am now saying that the functionalist analysis of worlds gives us a reason to reject the local branching view, I must

---

[21] In Ney (2025), I adopt perdurantism. However, see Romagosa (manuscript) for reasons to be cautious.

also now explain how it is that this argument is unsound. Recall the argument went as follows:

1. Branching is the result of decoherence.

2. Decoherence is a local causal process.

Therefore,

3. Branching is a local causal process.

My claim is that both premises of this argument are true. The first premise follows from the functionalist analysis of worlds and the definition of decoherent state. The second premise follows from the way that decoherence is typically achieved by environment-induced superselection. The problem with this argument is not therefore that it has a false premise, but that it is logically invalid. For note it is logically consistent to suppose that although branching is the result of decoherence and decoherence takes time, once the decoherence process is "finished," then branching occurs instantaneously and all at once. So, although decoherence is a local causal process, branching itself is not. This is what I propose we say. In the next section, we will iron out the details of how this can be a consistent way to interpret branching in a relativistic many worlds theory.

But first, it is worth noting the mistake that defenders of the local branching model make when they say branching is a local causal process. (This is most explicit in Blackshaw, Huggett and Ladyman this volume.) This is to conflate branching with decoherence. For the purposes of many discussions, this is harmless. But if we are trying to be precise about the metaphysical framework of the MWI, we should note that these two concepts apply to different kinds of phenomena. Decoherence is a microphysical process that is revealed in the quantum formalism, when formulated using density matrices. Branching is a concept that applies to emergent ontology on the basis of dispositional features that are tracked by counterfactuals. This leads us back to a point that was made in Section 2. There is a reason to understand fundamental physical processes as causal to avoid tensions with relativity. But this is less well-motivated in discussions of emergent ontology. More on this in a moment.

## 4. The Global Branching View is Not in Tension with Relativity

I have argued that the functionalist analysis of worlds accepted by most proponents of the MWI should lead us to adopt the view that branching is not a local causal process, but rather an instantaneous event taking place across the total entangled system at a time. In the extreme case in which there is global entanglement, this is across the entire universe at a time. The natural question to ask then is why this doesn't create a tension between the MWI and special relativity. There are two matters to address. The first is why global branching does not commit us to additional spatiotemporal structure in the form of preferred reference frames. The second is why the global branching view does not involve any superluminal influence.

Starting with the first matter, the global branching view is only committed to preferred reference frames if branching structure is absolute, that is, if there is an objective fact that could be agreed to by all observers about which objects have and have not branched at which times. However, recall, following Saunders (1993), Wallace (2010), and Carroll (2019), the branching structure is not a fundamental feature of a MWI metaphysics. As such, it need not be absolute. Although two observers, Alice and Bob, might agree that branching is global and happens everywhere and all at once, they need not agree which entities have branched at a given time and which have not. If they disagree on a reference frame, they will disagree which events are simultaneous with Alice's measurement, and so they will disagree about what has and has not branched.

What about superluminal influences? In Section 1, we introduced the question: does Alice's measurement immediately cause Bob to branch? According to the global branching view, the answer to this question is: yes. So, we may now ask, how is the proponent of the global branching view not committed to spooky action at a distance in the form of superluminal influence? The crucial point to note is that Alice's measurement has not resulted in any intrinsic, physical change to anything at Bob's location. The physical processes are best represented by the evolution from:

(1) $$\psi = \frac{1}{\sqrt{2}}|ready\rangle_A|ready\rangle_B|ready\rangle_{D_A}|ready\rangle_{D_B}|E_0\rangle_E\left(|\uparrow_z\rangle_a|\downarrow_z\rangle_b - |\downarrow_z\rangle_a|\uparrow_z\rangle_b\right)$$

to:

(4) $$\psi = \frac{1}{\sqrt{2}}|ready\rangle_B|ready\rangle_{D_B}\left(|\uparrow\rangle_A|\uparrow\rangle_{D_A}|\uparrow_z\rangle_a|\downarrow_z\rangle_b|E_1\rangle_E - |\downarrow\rangle_A|\downarrow\rangle_{D_A}|\downarrow_z\rangle_a|\uparrow_z\rangle_b|E_2\rangle_E\right).$$

As mentioned in Section 1, in the evolution from (1) to (4), there is no change to the reduced density matrix associated with Bob, his measuring device, or his particle. Assuming the plausible principle endorsed in Wallace and Timpson (2010) that intrinsic, physical features are best captured by reduced density matrices, it follows that there is no intrinsic physical change to anything at Bob's location.[22] Given the change to the circumstances at Alice's region, the physical situation at Bob's location now bears new relations to those at Alice's that will allow Alice to make true claims about branching. If Bob likes, he can make these claims as well, but he will do so knowing that he hasn't intrinsically changed in any way. In this way, it is just like your saying, "I now have a sister-in-law." You can note you now bear some interesting new relations to other objects in the world while at the same time recognizing that you haven't intrinsically changed in any way.[23]

---

[22] For an extensive defense of this claim and why branching is not an intrinsic change to an object, see Ney (2025) and also, independently, Faglia (2024).

[23] The point that changes to extrinsic (but not intrinsic) features don't make for any kind of troubling action at a distance is an old one in metaphysics. See, for example, Geach (1969) who introduced the distinction between real changes and mere Cambridge changes.

If we want to refer to branching as a process, then at best it is a pseudo-process, in the sense discussed by Wesley Salmon. Recall Salmon's example. A beacon of light is placed in the center of a circular room. As it rotates, it projects a spot of light on the wall. If the beacon rotates at a high enough (subluminal) rate and the room's radius is large enough, the spot may move around the room at a superluminal speed. This can produce a puzzle, until one recognizes that the real causal processes are just the light beams streaming from the beacon outward to the points on the wall:

> In the context of relativity theory, it is essential to distinguish causal processes, such as the propagation of a light ray, from various pseudo-processes, such as the motion of a spot of light cast on a wall by a rotating beacon… The motion of the spot is a well-defined process of some sort, but it is not a causal process. The causal processes involved are the passages of light rays from the beacon to the wall… This fact has great moment for special relativity, for the light beam can travel no faster than the universal constant *c*, while the spot can move across the wall at arbitrarily high velocities… The arbitrarily high velocities of pseudo-processes cannot be exploited to undermine the relativity of simultaneity. (Salmon 1975, p. 114)

The analogy with our case is not perfect. In Salmon's case there is an intrinsic change on the wall from one point to the next. In our case, when Alice conducts her measurement, there is not any intrinsic change where Bob is. Nonetheless, at the time Alice measures, we can correctly note (extrinsic, relational) differences that weren't there at Bob's location previously. We call this extrinsic change that takes place "branching." It is at best a pseudo-process.

## 5. Conclusion

In conclusion, one should not be concerned to interpret branching in the MWI as a global, instantaneous event. This is compatible with branching being triggered by decoherence, a local causal process that takes time. Branching in the MWI should thus be understood as a change that takes time to occur, since it depends on decoherence. However, once it occurs, it occurs everywhere and all at once, or at least everywhere across the total entangled system.

As we have seen, this view is motivated by the functionalist analysis of worlds. Once it is true for an Alice that were Bob (i.e. a Bob she is not causally isolated from) to measure his particle, he would find some definite result, then it is thereby true that there are multiple Bobs, one of which she is causally isolated from.

The global branching view does not create tension with special relativity. However, we must be careful to note that branching is only relative to an observer and a frame, and so observers may disagree about the time branching takes place. Since the branching "process" does not involve any physical changes to observers or other quantum systems, this undercuts any motivation to require it be a local process, as in Wallace (2012) and Blackshaw, Huggett, and Ladyman (this volume).

Global branching is the position that I have argued is most compatible with the functionalist analysis of worlds adopted by many advocates of the MWI. Worlds are generated globally and instantaneously once there is sufficient causal isolation between parts of the overall quantum state. However, it has been suggested to me that one might instead think there are two kinds of branching that play an important role in the MWI.[24] This leads to what may appear to be a more ecumenical position, one that allows the many worlds theorist to continue to apply the 'branching' label to the decoherence process. Call the process by which a coherent quantum state decoheres leading to causal isolation the *branching process*

[24] In conversation with Chris Timpson.

(or branching$_1$). This is the local process described by Wallace (2012) and Blackshaw, Huggett, and Ladyman (this volume). Call the instantaneous event that occurs when sufficient causal isolation has been achieved such that it is true to say that there are multiple worlds the *branching event* (or branching$_2$). I see no reason to reject this linguistic proposal. However, it is worth emphasizing that even if we agree to adopt it, there will remain a disagreement about how far across spacetime the branching event (branching$_2$) reaches. The argument I presented in Section 3.4 requires us to conclude that it reaches farther than advocates of the local branching view allow. Moreover, note that the many worlds theorist cannot make do with the decoherence process (branching$_1$) alone, as described in Blackshaw, Huggett, and Ladyman. For, as MWI advocates have been pressing for many years, decoherence alone does not solve the measurement problem. One needs the decoherence to successfully achieve a branching into worlds to achieve determinate measurement results.[25]

Work Cited

---

[25] Thanks to audiences at the University of Geneva, Oxford, the University of Cologne, and the 2024 Philosophy of Science Association Meeting in New Orleans. I especially want to thank Emily Adlam, David Albert, Jacob Barandes, Nadia Blackshaw, Sam Fletcher, Alex Franklin, Stephan Hartmann, Mario Hubert, Nick Huggett, James Ladyman, Siddarth Muthukrishnan, Paula Reichert, Kian Salimkhani, Chip Sebens, Chris Timpson, David Wallace, Al Wilson, and Jim Woodward for conversations that greatly improved the paper.